# Journal Name

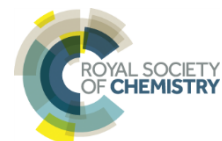

## ARTICLE

## Organic nanotubes created from mesogenic derivatives

Vladimíra Novotná,*[a] Věra Hamplová,[a] Lubor Lejček,[a] Damian Pociecha,[b] Martin Cigl,[a] Ladislav Fekete,[a] Milada Glogarová,[a] Lucie Bednárová,[c] Pawel W. Majewski[b] and Ewa Gorecka[b]

A facile route to prepare the nanotubes from rod-like mesogens dissolved in typical organic solvents is reported. For selected types of chiral rod-like molecules, both enantiomers as well as the racemic mixtures formed the nanotubes by slow evaporation from a solution, regardless of the solvent, concentration or deposition type. Obtained supramolecular assemblies were studied using AFM and SEM techniques. Additional methods (IR, UV-Vis spectroscopy and x-ray diffraction) were applied to characterize the observed nanotubes. The difference in the surface tension at opposite crystallite surfaces is suggested as a possible mechanism for the nanotube nucleation.

## Introduction

Nanotubular morphology can be observed for various materials. The most studied examples are carbon nanotubes (CNTs) formed by rolling up graphite sheets around the edges.[1] Since the discovery of CNTs[2] other types of layered inorganic nanotubes have been described, e.g. metal chalcogenides $WS_2$, $MoS_2$ or $TiO_2$ nanotubes.[3,4] Likewise, nanotubes can be built from organic molecules and macromolecules, the organic nanotubes (ONTs) are of a great interest and represent a challenging topic in the area of materials chemistry. Various types of ONTs have been synthesized and their applications actively explored in recent years.[5] They offer potential applications as electronically and biologically active materials for their better biodegradability and biocompatibility in comparison with inorganic materials with the covalent linkages.[6] In addition, ONTs' surfaces, morphologies, sizes, and functionalities are easily tuned by modification of the nanotube-forming molecules or the self-assembly process. The utilization of the inner channel of the nanotubes has also attracted much attention as it offers a potential for filtration applications[7] and a possibility to capture and release various guest molecules in a controllable way.[8] Various methods to control ONTs' outer surfaces for functionalization, manipulation, and intermolecular organization have also been developed.[9]

Organic nanotubes spontaneously self-assemble through various molecular interactions such as a hydrogen bonding, van der Waals forces, and π-π stacking; molecular shape and other conditions are playing a role as well.[5] As a driving force for the scrolling of flat sheets into nanotubes, the chirality of the materials or the layer curvature was suggested,[10] or the polarization charges at the opposite sides of the sheet.[3] Reducing the surface energy of a flat sheet in solution is also a possible scrolling mechanism that leads to the formation of ONTs.[5] The largest group of materials forming ONTs are the amphiphilic molecules possessing hydrophilic and hydrophobic parts and a few reviews have been published recently.[10,11] In most cases the lipid nanotubes involve the helically-coiled ribbon structures.[12,13] Another sort of amphiphilic compounds is represented by the diamides.[13] The nanotubes derived from the terphenylene-diamide have been proved to exhibit a scroll-type structure formed by the rolled-up sheets.[14] Only rarely the ONTs are built from the molecules with none or small asymmetry in molecular structure.[15]

Liquid crystalline (LC) phases are formed by the self-assembly of organic molecules; most commonly observed are mesophases with lamellar morphology (smectics). In rare cases, the morphology of such phases is more complex than just the flat layers e.g. nano-ribbons or sponge type structures can be formed.[16,17,3a] So far, a formation of nanotubes for mesogenic compounds has been reported only for bent-core molecules exhibiting B4 phase.[18]

Herein, we present a new type of mesogenic rod-like compounds, which spontaneously form nanotubes in a crystalline state. The studied chiral rod-like molecules are in the crystalline state at the room temperature and exhibit a sequence of liquid crystalline phases above the melting point.[19] The rolled-up nanotubes are created from a solution during the solvent evaporation. The nanotube formation process was studied as a function of the molecular structure and the type of the solvent. We used atomic force microscopy

[a.] Institute of Physics, Czech Academy of Sciences, Na Slovance 2, CZ-182 21 Prague 8, Czech Republic. E-mail: novotna@fzu.cz; Fax: +420286890527; Tel: +420266053111.
[b.] Faculty of Chemistry, University of Warsaw, ul. Zwirki i Wigury 101, 02-089 Warsaw, Poland. E-mail: gorecka@chem.uw.edu.pl; Tel: +48228221075.
[c.] Institute of Organic Chemistry and Biochemistry of the CAS, Flemingovo n. 2,166 10 Prague 6 CZ-182 21 Prague 6, Czech Republic.

Electronic Supplementary Information (ESI) available: [details of any supplementary information available should be included here]. See DOI: 10.1039/x0xx00000x

(AFM) and electron microscopy (SEM) to study the morphology of the nano-assemblies. The infrared and UV spectroscopy and the x-ray measurements provided additional information about the self-assembly behaviour at the molecular level.

## Results and discussion

Chemical formulas of the studied compounds, denoted as 10ZBBL, 10ZBDL and mZBL, are shown in Fig. 1. Materials mZBL have been prepared by a synthetic route described previously,[19] the details of 10ZBBL and 10ZBDL syntheses are given in ESI. Compounds 10ZBBL and 10ZBDL have been synthesized in both enantiomeric forms and the racemic mixtures were also prepared. At room temperature all studied compounds are crystalline and melt to the LC smectic phase at about 65ºC. The mesomorphic properties of mZBL, which contains two S- type chiral centres were studied previously, the compound exhibited unusual sequence of smectic phases.[19]

The nanotubes were prepared at the room temperature, the molecules were dissolved in an organic solvent and the solution was spread on a solid surface, the subsequent slow spontaneous evaporation of the solvent yielded thin films. The morphology of the films has been studied using AFM and SEM techniques. The methods (AFM, SEM, IR, UV-Vis spectroscopy and x-ray diffraction) and equipment used in the reported study are described in details in ESI.

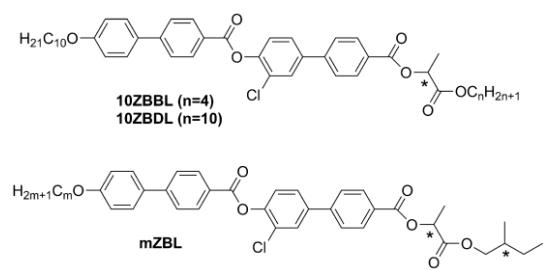

**Fig. 1.** Chemical formula of 10ZBBL, 10ZBDL and mZBL compounds.

The nanotube growth was tested in details under different conditions for 10ZBBL and mZBL materials (see Fig. 1). We checked several types of organic polar or non-polar solvents, in which the compounds are soluble: cyclohexane, toluene and acetone, and various concentrations: $10^{-3}$ - $10^{-2}$ mol/l. Also, the influence of the solvent evaporation rate was verified. Additionally, the nanotubes were prepared on several types of solid surfaces: glass, KBr, sapphire and lithium niobate (LiNbO$_3$) and it was established that the nanotubes growth from the solution does not depend on the type of the substrate. For 10ZBBL and mZBL materials the shape and size of the obtained nanotubes were very similar. No differences were found while using pure enantiomers or racemic mixture of 10ZBBL, thus we can exclude the optical purity of the compound as a necessary condition for the nanotubes' growth. The AFM images revealed that the nanotube diameter is 50 to 60 nm with the inner channel diameter of ~ 20 nm (Fig. 2), regardless of the film preparation conditions. On the other hand, the length of nanotubes varied from sample to sample. Close inspection of the nanotube's surface evidenced a roll-up mechanism of their formation. The thickness of the nanotube wall is ~15 nm and small 3-5 nm steps are visible. Thus the nanotube wall comprises around 4 mono-molecular layers wound one over another. Apart of the typical organic solvents, we have also tested a chiral solvent (*S*)-(2-methylbutyl)methylether (see SI for more information) and confirmed the nanotubes' presence for both pure enantiomers and racemic mixture of 10ZBBL, no chiral discrimination effects were observed.

The AFM images showed that in majority the nanotubes grow parallel to the surface (Figs. 2,3), only in few images we were able to detect the nanotubes oriented with their axes perpendicular to the surface (inset in Fig. 2, Fig. S3 in ESI). Moreover, in the numerous experiments the tubes were found mutually parallel to each other over the large areas (Fig. 3). Under a weak shearing between glass plates or polymer foils we were able to orient the nanotubes in macroscopic areas, with their axes parallel to the shearing direction, which was proved by the x-ray diffraction studies (Fig. 4). The high quality of the alignment was tested by performing x-ray diffraction from the different sample areas.

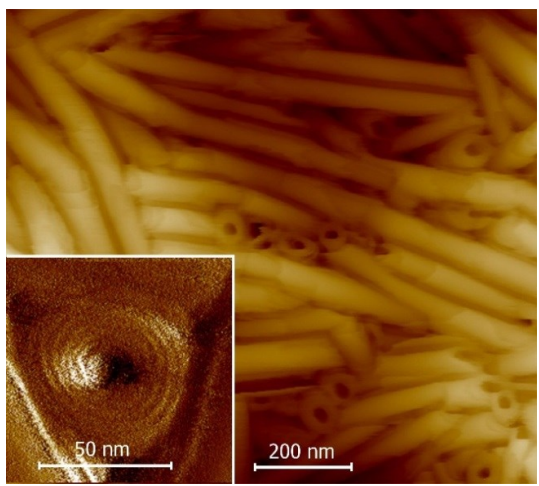

**Fig. 2.** AFM images of 10ZBL film grown by the evaporation from acetone. In the inset the enlarged view of the nanotube oriented perpendicular to the surface is shown.

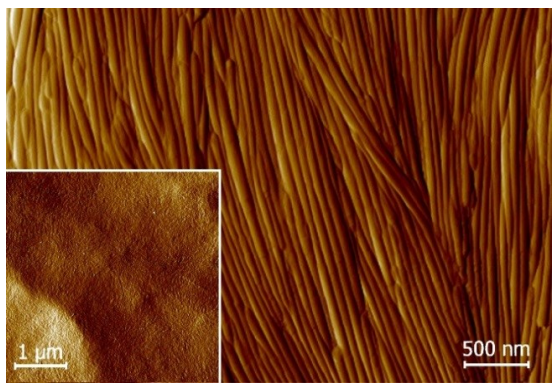

**Fig. 3.** AFM image of 10ZBL film on LiNbO$_3$ substrate shows a parallel alignment of the nanotubes over large area. In the inset the AFM image of the crystalline phase of 10ZBBL(rac) obtained for a sample recrystallized from the smectic phase and a smooth surface with no complex morphology was observed.

The x-ray studies confirmed the crystalline character of the nanotubes (Fig. 4); the large width of the recorded x-ray diffraction signals was a result of a highly-limited size of the diffracting objects.[20] The material crystallized in orthorhombic unit cell (for 9ZBL the crystallographic unit cell parameters are $a$=42 Å, $b$=15 Å and $c$=4.5 Å). The crystal had a lamellar structure with the layer spacing ($a$ crystallographic parameter) similar to that found in the smectic phase and which corresponded to the molecular length. The orientation of the signals in the XRD patterns for sample aligned by shearing suggested that the nanotube axis was oriented along the $b$ crystallographic unit cell direction. The nanotubes melted at the transition from the crystalline to the liquid crystalline phase (SmC). On cooling back, the material recrystallized in the same crystalline form as found for ONTs (Fig. 4), but neither nanotubes nor any other supramolecular morphology was observed (Fig. 3 inset); apparently the presence of the solvent is crucial for the nanotube formation.

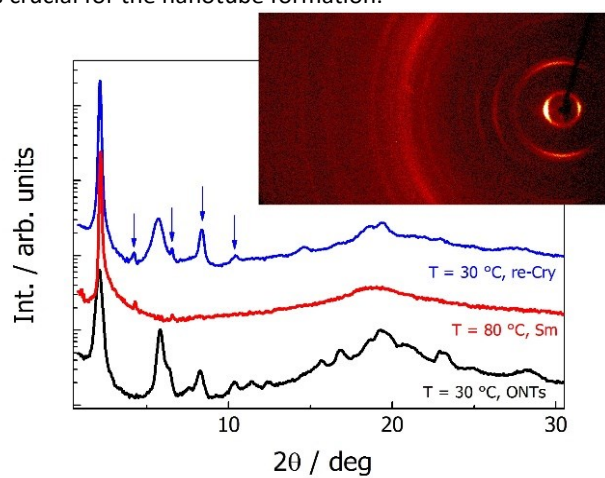

**Fig. 4.** X-ray diffraction patterns of compound 9ZBL(S): the nanotubes obtained from the toluene (black line), the smectic phase at T=80 °C (red line) and crystal formed from the melt (blue line). Arrows indicate harmonics of the main signal related to the layer thickness. In the inset the 2D pattern obtained for ONTs is shown.

The ONTs on KBr substrate were studied using IR spectroscopy from the room temperature until the material melted to the smectic phase and the spectra (3600 – 600 cm$^{-1}$) are presented in ESI (Figs. S9-S10). The main difference between the ONTs and liquid crystalline phase was found in the C-H stretching spectral region (3000 - 2800 cm$^{-1}$). The spectral shift of the absorption maxima from 2922 cm$^{-1}$ to 2927 cm$^{-1}$ and 2852 cm$^{-1}$ to 2856 cm$^{-1}$ assigned to $\nu_A$(C-H) and $\nu_S$(C-H) of CH$_2$ groups, respectively, shows that the aliphatic chains liquidise at the transition to the liquid crystalline state.[21] The spectral shifts were also observed for the absorption maxima related to the carbonyl groups, for ONTs at 1724 cm$^{-1}$ (ester group at a terminal phenyl ring), 1744 cm$^{-1}$ (ester group between biphenyl units) and 1752 cm$^{-1}$ (ester group in the chiral tail). In the smectic phase the first two moved to 1726 cm$^{-1}$ and 1739 cm$^{-1}$, respectively, with comparable intensity as at room temperature. The intensity of band at 1752 cm$^{-1}$ decreases, probably correspondingly to its degree of freedom,[22] and shifts to 1760 cm$^{-1}$. In the crystalline state of the ONTs the carbonyl group in the mesogenic core violation of co-planar with phenyl rings arrangements it is not excluded considering the spectral change of carbonyl and spectral shift and intensity change of the band at 1602 cm$^{-1}$, which is assigned to phenyl ring vibration. Apparently, the transition from ONTs to the LC state is accompanied by the molecular conformational change. The creation of nanotubes causes the shift the absorption maxima in the UV-Vis spectral range towards the shorter wavelength in comparison to its reference position for the non-interacting molecules in solution (Fig. S11). Such a hypsochromic shift indicates the presence of H-aggregates in crystalline structure with the parallel orientation of the neighbouring molecules.

We have also analysed the films prepared from 10ZBDL compound having longer chiral alkyl chain (n=10) than 10ZBBL compound (n=4) and did not observed formation of nanotubes. The AFM pictures for 10ZBDL showed only flat 3-4 nm thick crystal plates (Fig. S5). Moreover, among the homologues mZBL with various length of the non-chiral alkyl chain [19] (m = 5-10) only compounds m=8, 9 and 10 were found to form the nanotubes. This clearly indicates a very delicate balance of forces responsible for the nanotube creation with respect to the molecular structure.

The obtained results show that the mechanism responsible for the crystal layer curving and thus nucleation of the nanotubes is related to the crystal-solvent interactions that are different at the opposite surfaces of the crystal layer. The broken up-down symmetry of the molecular orientation, which creates different surfaces of the crystal layer or/and the evaporation process itself, which leads to the different solvent concentration at the opposite crystal surfaces, might play an important role. Irrespective of the molecular packing in layers, the process of the nanotube formation can be described using a phenomenological parameter - "surface tension". It is clear that the lack of balance in surface tension between two sides of a molecular layer becomes a driving force for the formation of a curved object, in the simplest case a rolled-up sheet. Let us determine the curvature radius, $r$, of a growing layer. The length of the layer arc of angle $\alpha$ is $r\alpha$. Supposing that the growth of curved layer is going on under a statistical equilibrium, the tangential force (per unit length in the direction parallel to the axis of layer rotation) on the inner surface, $-\partial(\gamma_1 r\alpha)/r\partial\alpha$, should be balanced by the tangential force on the outer surface, $-\partial(\gamma_2 r\alpha)/r\partial\alpha$, where $\gamma_1$ and $\gamma_2$ are the surface tensions of the inner and outer surfaces of a curved layer, respectively. The density of the elastic energy for the curved layer can be written[23] as:

$$f_{el} = \frac{K_1}{2}\left(\frac{1}{r_1} + \frac{1}{r_2}\right)^2 + \frac{K_2}{r_1 r_2}, \quad (1)$$

where $K_1$ and $K_2$ are the elastic constants and $r_1$ and $r_2$ are the principal radii of curvature of the layer. For cylinders, $r_1 = r$ and $r_2 = \infty$, thus the elastic energy density of a curved layer reduces to $\frac{K_1}{2}\left(\frac{1}{r}\right)^2$. The volume of the single crystal layer (per unit length) is $hr\alpha$, where $h$ is the layer thickness. Then, the tangential force due to the curvature of a layer (per unit length) will be given approximately by $-\partial(\frac{K_1}{2}\frac{h\alpha}{r})/r\partial\alpha$. Finally, we obtain the force balance for a growing curved layer (per unit length along the rotation axis[24] (see Fig. 5) in the form:

$$\gamma_1 = \gamma_2 + \frac{K_1 h}{2r^2} \quad (2)$$

which gives the radius of the curved layer:

$$r = \sqrt{\frac{hK_1}{2(\gamma_1-\gamma_2)}} \quad (3)$$

This approximation is valid for $r > h$. The radius $r$ exists if $\gamma_2 < \gamma_1$ and for $\gamma_1 \to \gamma_2$ the radius increases. Therefore in this model the curvature of a layer is caused by the difference of the surface energies on both layer surfaces; for equal surface energies no curvature of a layer occurs.

So far we considered a single layer of a molecular thickness. However, if $n$ layers are assembled together making a wall of the thickness $nh$, then the energy of a nanotube (per unit length) can be rewritten as:

$$\pi K_1 ln\frac{R}{r} + 2\pi(\gamma_1 r + \gamma_2 R). \quad (4)$$

In (4) the nanotube was considered in the form of a cylinder with the inner radius $r$ and outer radius $R = r + nh$. The surface energies contribute to the nanotube energy on the inner and outer surfaces only. The lowest energy of the nanotube corresponds to $n=1$, which supports our assumption that the nanotube nucleation starts from a single molecular layer, which curves in contact with the solvent due to the surface energy differences. Observations show that the inner nanotube radius is typically about 10 nm. Assuming that the single crystal layer in a solvent has similar bend elastic constant as a smectic liquid crystal, $K_1 \approx 10^{-11}$ J/m, for a layer of thickness $h \approx 3.5$ nm we can estimate $\Delta\gamma = \gamma_1 - \gamma_2$ from eq. (3) as $\Delta\gamma \approx 1.75 \times 10^{-4}$ J/m$^2$, that is only ~1% of surface tension of alkyl esters.[25] Such a small $\Delta\gamma$ difference can easily result from the broken up-down symmetry of the molecular orientation within crystal layer or from the solvent concentration gradient across the sample thickness.

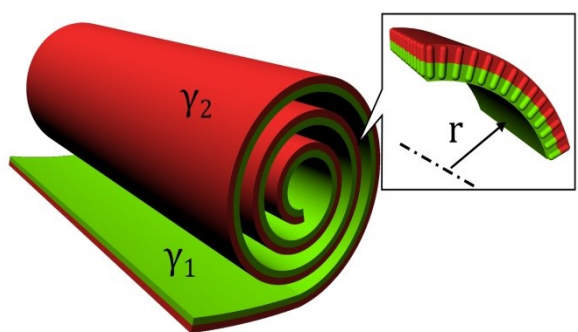

**Fig. 5.** Scheme of layers' rolling up and the ONT growth, coefficient $\gamma$ describes different surface tension on the opposite side of the layer. In the inset molecules' orientation is shown schematically.

## Conclusions

To conclude, let us point out that the character of the studied compounds is rather unusual for the ONTs as the molecules are not amphiphilic, having alkyl chains at both molecular ends. The nanotubes' morphology is not influenced by the type of the organic solvent and type of a substrate at which they are grown. Nanotubes formed by 10ZBBL compound were observed for both R and S enantiomers as well as for the racemic mixture, in achiral and chiral solvents, therefore we can exclude the chiral discrimination effects for nanotube formation. On the other hand, the molecular end group modification easily leads to the loss of the nanotube growth when the length of both terminal chains becomes similar. The molecules forming ONTs have different terminal parts and thus asymmetry of inner and outer surface of ONTs occurs. This supports the hypothesis that the main factor governing the nanotube formation is the asymmetry in surface tension at the opposite surfaces of the crystal layer, resulting from broken up-down symmetry of the molecular arrangement within the crystal layer. In spite of the fact we are not able to tune the size of nanotubes, we have succeeded in the nanotube alignment in micron size areas by shearing.

## Conflicts of interest

There are no conflicts to declare.

## Acknowledgements

This research was supported by project 18-14497S (the Czech Science Foundation). LF acknowledges grant MEYS LO1409 Infrastructure SAFMAT LM2015088; EG, DP and PWM thank for the support received from the National Science Centre (Poland) under Grant No. 2015/19/P/ST5/03813 (Polonez).